\documentclass[%
 reprint,
 amsmath,amssymb,
 aps,
 prd,
 floats,
]{revtex4-1}

\usepackage{graphicx}
\usepackage{dcolumn}
\usepackage{bm}

\begin{document}

\title{Cosmic Acceleration Caused by the Extra-Dimensional Evolution in a Generalized Randall-Sundrum Model}

\author{Guang-Zhen Kang}
\email{gzkang@nju.edu.cn}
\affiliation{School of Science, Yangzhou Polytechnic Institute, Yangzhou 225127, China}
\affiliation{Department of Physics, Nanjing University, Nanjing 210093, China}

\author{De-Sheng Zhang}
\affiliation{School of Science, Changzhou Institute of Technology, Changzhou 213032, China}

\author{Long Du}
\affiliation{School of Science, Guangxi University of Science and Technology, Liuzhou 545006, China}

\author{Dan Shan}
\affiliation{School of Science, Yangzhou Polytechnic Institute, Yangzhou 225127, China}
\affiliation{Department of Physics, Nanjing University, Nanjing 210093, China}

\author{Hong-Shi Zong}
\affiliation{Department of Physics, Nanjing University, Nanjing 210093, China}
\affiliation{Joint Center for Particle, Nuclear Physics and Cosmology, Nanjing 210093, China}
\affiliation{State Key Laboratory of Theoretical Physics, Institute of Theoretical Physics, CAS, Beijing 100190, China}



\date{\today}

\begin{abstract}
We investigate a $(n+1)$-dimensional generalized Randall-Sundrum model with an anisotropic metric which has three different scale factors.
One obtain a positive effective cosmological constant $\Omega_{eff}\sim10^{-124}$ (in Planck unit) which only need a solution $kr\simeq50-80$ without fine tuning,
and both the visible and hidden brane tensions are positive which results in the two branes to be stable.
Then, we find that the Hubble parameter is seem to be a constant in a large region near its minimum, thus causing the acceleration
of the universe.
Therefore, the fine tuning problem also can be solved in this model.
Meanwhile, the scale of extra dimensions is smaller than the observed scale but greater than the Planck length.
This demonstrates that the observed present acceleration of the universe is caused by the extra-dimensional evolution rather than dark energy.
\begin{description}
\item[PACS numbers]
04.50.-h, 11.25.Mj, 98.80.Es
\end{description}
\end{abstract}

\maketitle


\section{Introduction}

The current cosmic acceleration is an unexpected picture of the universe, revealed by the data sets of last two decades from astrophysics and cosmology \cite{Riess,Perlmutter,Bennett,Netterfield,Halverson,Valent,Zhang,Verde,Simon,Moresco1,Moresco2,Ratsimbazafy,Stern,Moresco3}.
These data, which coming from the cosmic microwave background radiation, supernovae surveys and baryon acoustic oscillations, etc, indicate that the universe consists of 4\% ordinary baryonic matter, 20\% dark matter and 76\% dark energy \cite{Peebles}.
Dark energy not only has an unknown form of energy but also has not been detected directly.
Additionally, dark energy is very similar to the cosmological constant which was proposed by Einstein.
In Planck unit, the observed value of the cosmological constant is an extravagantly tiny positive value of order $10^{-124}$.
This is the well-known cosmological constant fine tuning problems \cite{Amendola}.
There have been numerous attempts in order to solve this problem, such as quintessence, anthropic principle, $f(R)$ model, etc. \cite{Peebles88,Wetterich,Zlatev,Caldwell,Feng,Sotiriou,Dvali,Bousso}.
But none of these theories are problem-free.
In astrophysics and cosmology, it is still a very important problem.

Another perspective for resolving the above described problem, which seems to be more radical, is the following:
The dimensions of our universe must be four? Are there some extra dimensions which are too small to be observed?
Does the evolving of these extra dimensions contribute to the current cosmic acceleration?
If so, would this help in solving the cosmological constant fine tuning problem?
Therefore, we have investigated some higher-dimensional theories \cite{KK,NAH1,NAH2,RS,Das,Antoniadis,Das1,Sundrum,Lykken,Antoniadis1,Visinelli,Vagnozzi,Paul,Polchinski}.
Among them, the Randall-Sundrum (RS) two-brane model \cite{RS},
which has a natural solution to the hierarchy problem with warped extra dimension,
has attracted our attention.
The hierarchy problem is essentially a fine tuning problem which can be described as: why there is such a large discrepancy between the electroweak scale/Higgs mass $M_{EW}\sim1$TeV and the Planck mass $M_{pl}\sim10^{16}$TeV?
In RS two-brane scenario, our universe is described by a five dimensional line element \cite{RS}
\begin{eqnarray}
ds^{2}=e^{-2\sigma(y)}\eta_{\mu\nu}dx^{\mu}dx^{\nu}+r_{c}^{2}dy^{2},
\end{eqnarray}
where $y$ is the extra dimensional coordinate, $r_{c}$ is the extra dimensional compactification radius,
$e^{-2\sigma}$ is the well-known warp factor, the term $\sigma=kr_{c}|\phi|$,
$k=\sqrt{-\Lambda/24M^{3}}$, $M$ is the five dimensional Planck mass.
Then a large hierarchy is generated by the warp factor $e^{-2kr_{c}\pi}$, meanwhile one requires only $kr\approx10$.
The cosmological constant fine tuning problem is similar to the hierarchy problem.
In RS model, the visible brane is instable caused by the negative brane tension.
Furthermore, the cosmological constant on the visible brane is zero which is not consistent with our data sets of last two decades \cite{Das,Koley}.

The above problems can be solved in a generalized RS braneworld scenario in which $g_{\mu\nu}$ replaces $\eta_{\mu\nu}$ in RS model \cite{Das}.
In this scenario, the tension of the visible brane and the hidden brane can be both positive with a negative induced cosmological constant.
It is very interesting because that both branes are stable \cite{Mitra,SC1,SC2,SC3,Banerjee,Kang1}.
In order to be consistent with the current constraints,
the negative induced cosmological constant $\Omega$ should be transformed into the positive effective cosmological constant $\Omega_{eff}$.
This positive $\Omega_{eff}$ can be obtained in a $(n+1)$-dimensional (-d) generalized RS model with two $(n-1)$-branes instead of two 3-branes \cite{Kang2}.
In this model, adopting an anisotropic metric ansatz with two different scale factors,
one obtain the positive effective cosmological constant $\Omega_{eff}\sim10^{-124}$ (in Planck unit) which only need a solution $kr\simeq50-80$ without fine tuning.
The cosmological constant fine tuning problem can be solved quite well \cite{Kang2}.

But there is no reason to exclude the possibility of the anisotropic metric ansatz with the form of scale factors more than two.
In this paper, we investigate a $(n+1)$-dimensional generalized Randall-Sundrum model with an anisotropic metric which has three different scale factors.
We obtain that $H_{1}$ has a lower bound $H_{1min}$.
Near this minimum value, the Hubble parameter is seem to be a constant in a large region, thus causing the acceleration of the universe.
Meanwhile, the scale of extra dimension is smaller than the observed scale but greater than the Planck length.
This demonstrates that the observed present acceleration of the universe is caused by the extra-dimensional evolution rather than dark energy.
Our work is organized as follows: In Sec. \uppercase\expandafter{\romannumeral2},
by considering the two $(n-1)$-branes with the matter field Lagrangian in $(n+1)$-d generalized RS model,
the $n$-d Einstein field equations are obtained.
In Sec. \uppercase\expandafter{\romannumeral3},
we focus on the evolution of $(n+1)$-brane solved from the above field equation with an anisotropic metric ansatz which has three different scale factors.
Finally, the summary and conclusion are presented in Sec. \uppercase\expandafter{\romannumeral4}.

\section{$(n+1)$-d Generalized Randall-Sundrum Model}
We consider a $(n+1)$-d generalized RS braneworld model which is consistent with the Ref.\cite{Kang2}.
The action $S_{n+1}$ is:
\begin{eqnarray}
S_{n+1}=S_{bulk}+S_{vis}+S_{hid} \label{eq:S},
\end{eqnarray}
where $S_{bulk}$ is the bulk action,
$S_{vis}$ and $S_{hid}$ are the $(n-1)$-brane visible action and hidden action, respectively:
\begin{eqnarray}
S_{bulk}&=&\int d^{n}xdy\sqrt{-G}(M^{n-1}_{n+1}R-\Lambda)\label{eq:S1},\\
S_{vis}&=&\int d^{n}x\sqrt{-g_{vis}}(\mathcal{L}_{vis}-V_{vis}),\\
S_{hid}&=&\int d^{n}x\sqrt{-g_{hid}}(\mathcal{L}_{hid}-V_{hid}),
\label{eq:S2}
\end{eqnarray}
where $\Lambda$ denotes a bulk cosmological constant,
$M_{n+1}$ is the $(n+1)$-d fundamental mass scale,
$G_{AB}$ and $R$ are the $(n+1)$-d metric tensor and Ricci scalar respectively,
$\mathcal{L}_{i}$ is the matter field Lagrangian of the visible and hidden branes,
$V_{i}$ is the tension of the visible and hidden branes, here $i=hid$ or $vis$.
In this $(n+1)$-d generalized RS scenario, the metric takes the form:
\begin{eqnarray}\label{eq:ds2}
ds^{2}=G_{AB}dx^{A}dx^{B}=e^{-2A(y)}g_{ab}dx^{a}dx^{b}+r^{2}dy^{2}.
\end{eqnarray}
where $e^{-2A(y)}$ is known as the warp factor,
Capital Latin $A,B,...$ indices run over all spacetime coordinate labels,
$y$ is the extra dimensional coordinate of length $r$,
Lowercase Latin $a,b=0,1,2,\cdot\cdot\cdot,n-1$ which is not include the coordinate $y$,
$g_{ab}$ is the $n$-d metric tensor.
Variation with respect to the metric $G_{AB}$ and after some easy manipulations, then modulo surface terms, one obtains
\begin{eqnarray}\label{eq:field}
R_{AB}-\frac{1}{2}G_{AB}R=\frac{1}{2M^{n-1}_{n+1}}\{-G_{AB}\Lambda+\sum_{i}[T^{i}_{AB}  \nonumber\\
\times\delta(y-y_{i})-G_{ab}\delta_{A}^{a}\delta_{B}^{b}V_{i}\delta(y-y_{i})]\},
\end{eqnarray}
where $R_{AB}$ is the $(n+1)$-d Ricci tensor, $T^{i}_{AB}$ is the $(n+1)$-d energy-momentum tensors.
Note here the energy-momentum tensor is given by $T^{ia}_{b}=diag[-c_{i},c_{i},\cdot\cdot\cdot,c_{i}]$ \cite{Kang1,Kang2},
A solution to the Eq.~(\ref{eq:field}) with the metric tensor Eq.~(\ref{eq:ds2}) has been derived in Ref.~\cite{Kang2}, which reads
\begin{eqnarray}\label{eq:solution-}
A=-\ln[\omega\cosh(k|y|+c_{-})],
\end{eqnarray}
where the constant $k\equiv\sqrt{-\Lambda/[M^{n-1}_{n+1}n(n-1)]}\simeq$ Planck mass, $\omega$ is given by
\begin{eqnarray}
\omega\equiv\sqrt{\frac{-2\Omega}{(n-1)(n-2)k^{2}}},
\end{eqnarray}
the term $c_{-}$ takes the form
\begin{eqnarray}
c_{-}\equiv\ln\frac{1-\sqrt{1-\omega^{2}}}{\omega}.
\end{eqnarray}
Meanwhile, a $n$-d Einstein field equations can be obtained
\begin{eqnarray}\
\widetilde{R}_{ab}-\frac{1}{2}g_{ab}\widetilde{R}=-\Omega g_{ab}. \label{eq:ndfield}
\end{eqnarray}
where $\Omega$ is the induced cosmological constant on the visible brane,
$\widetilde{R}$ and $\widetilde{R}_{ab}$ are the $n$-d Ricci scalar and Ricci tensor respectively,

Note that the solution derived above has the negative induced cosmological constant $\Omega$,
Here we do not consider the situation for $\Omega>0$,
since the tension on the visible brane is negative which results in instability \cite{Das,Koley,Kang1,Kang2}.
Then, an anisotropic metric is assumed to be of the following form \cite{Middleton,Kang1,Kang2}:
\begin{eqnarray}
g_{ab}=diag[-1,a_{1}^{2}(t),a_{2}^{2}(t),a_{3}^{2}(t),\cdot\cdot\cdot,a_{n-1}^{2}(t)], \label{eq:gab}
\end{eqnarray}
where $a_{i}$ is the scale factor.
The case that the scale factors on the visible brane evolve with two different rates has been studied recently \cite{Kang2}.
In this case, we can obtain the positive effective cosmological constant $\Omega_{eff}\simeq{10}^{-124}$ and only requiring $kr\simeq50-80$£¬
where for convenience, the Planck mass has been set to unity.
So the cosmological constant fine tuning problem can be solved quite well.
Furthermore, the three dimensional (3D) Hubble parameter $H(z)$ is consistent with cosmic chronometers dataset extracted from \cite{Valent,Zhang,Verde,Moresco1,Moresco2,Moresco3,Ratsimbazafy,Stern,Simon}.
The observed 3D universe naturally shifts from deceleration expansion to accelerated expansion.
It shown that the accelerated expansion of the observed universe is intrinsically an extra-dimensional phenomenon.
But, there is no reason to make the scale factor evolve only with two kinds of rates.
Therefore, we investigate the case that the scale factors on the visible brane evolve with three different rates.

\begin{figure*}
\includegraphics[scale=0.85]{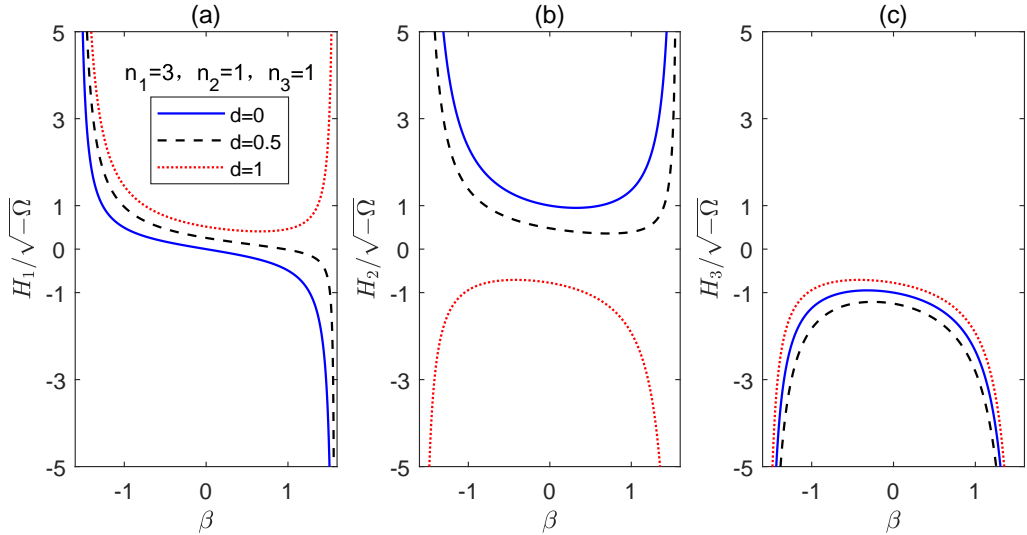}
\caption{\label{fig:1} The Hubble parameters $H_{1}$ ($n_{1}=3$), $H_{2}$ ($n_{2}=1$) and $H_{3}$ ($n_{3}=1$) with three different parameters $d$.
(a) The Hubble parameters $H_{1}$ with $d=0, 0.5, 1$ correspond to the solid (blue) curve, the dashed (black) curve and the dotted curve (red) respectively ($n_{2}=1$). (b)-(c) The case for the Hubble parameters $H_{2}$ and $H_{3}$ respectively.}
\end{figure*}

\section{Anisotropic Evolution of $(n-1)$-Brane}

For the anisotropic metric Eq.~(\ref{eq:gab}) with three kinds of scale factors
and the negative induced cosmological constant $\Omega\sim-10^{-124}$,
the field equations~(\ref{eq:ndfield}) can be written:
\begin{eqnarray}
\sum_{i}n_{i}(n_{i}-1)H_{i}^{2}+\sum_{i\neq j}n_{i}n_{j}H_{i}H_{j}=2\Omega, \label{eq:field1}\\
\sum_{i}n_{i}\dot{H}_{i}-\dot{H}_{1}+(\sum_{i}n_{i}H_{i})^{2}-H_{1}\sum_{i}n_{i}H_{i}=2\Omega, \label{eq:field2}\\
\sum_{i}n_{i}\dot{H}_{i}-\dot{H}_{2}+(\sum_{i}n_{i}H_{i})^{2}-H_{2}\sum_{i}n_{i}H_{i}=2\Omega, \label{eq:field3}\\
\sum_{i}n_{i}\dot{H}_{i}-\dot{H}_{3}+(\sum_{i}n_{i}H_{i})^{2}-H_{3}\sum_{i}n_{i}H_{i}=2\Omega, \label{eq:field4}
\end{eqnarray}
where $i=1,2,3$, the terms $n_{1}$, $n_{2}$ and $n_{3}$ are the number of dimensions which evolve with three kinds of rates respectively,
the Hubble parameter $H\equiv\dot{a}/a$, $\dot{H}_{i}$ is the first time derivative of ${H}_{i}$.
Computing the sum of Eqs.~(\ref{eq:field2}), (\ref{eq:field3}) and (\ref{eq:field4}),
yields a simplified expression for $\sum_{i}n_{i}H_{i}$:
\begin{eqnarray}
\sum_{i}n_{i}H_{i}=-\chi_{1}\tan\beta, \label{eq:sumH}
\end{eqnarray}
the term $\beta=\chi_{1}t+\theta_{0}$,
$\theta_{0}$ is the initial phase angle which is determined by the scale of the formation of the brane,
the terms $\chi_{1}$ takes the form
\begin{eqnarray}
\chi_{1}=\sqrt{\frac{-2(n-1)\Omega}{n-2}}.\label{eq:chi1}
\end{eqnarray}
It is convenient to redefine the sum of the Hubble parameters in the following
\begin{eqnarray}
\sum_{i}n_{i}H_{i}\equiv f.\label{eq:f}
\end{eqnarray}
Replacing Eqs.~(\ref{eq:sumH}) and (\ref{eq:f}) in Eq.~(\ref{eq:field2}), then after some easy manipulations, one obtains
\begin{eqnarray}
\dot{H}_{1}+H_{1}f=\dot{f}+f^{2}-2\Omega.\label{eq:H1eq}
\end{eqnarray}
The solution of the above equation is
\begin{eqnarray}
H_{1}=-\frac{\chi_{1}}{n-1}\tan\beta+c\sec\beta\label{eq:3h1},
\end{eqnarray}
where $c$ is an integration constant.
Eq.~(\ref{eq:field1}) can be rewritten as
\begin{eqnarray}
n_{2}H_{2}^{2}+n_{3}H_{3}^{2}=-n_{1}H_{1}^{2}+f^{2}-2\Omega.   \label{eq:3h2h3}
\end{eqnarray}
The Eqs.~(\ref{eq:3h1}) and (\ref{eq:3h2h3}) could be combined to give the following equation eliminating $H_{2}$ completely
\begin{eqnarray}
(n_{3}+\frac{n_{3}^{2}}{n_{2}})H_{3}^{2}+\frac{2n_{3}}{n_{2}}(n_{1}H_{1}-f)H_{3}+[(\frac{1}{n_{2}}-1)f^2 \nonumber\\
+\frac{n_{1}^{2}}{n_{2}}H_{1}^{2}-\frac{2n_{1}}{n_{2}}fH_{1}+n_{1}H_{1}^{2}+2\Omega]=0.  \label{eq:H3eq}
\end{eqnarray}
Then the Hubble parameter $H_{3}$ can be obtained
\begin{eqnarray}
H_{3}=-\frac{\chi_{1}}{n-1}\tan\beta-\frac{\chi_{3}+n_{1}c}{n_{2}+n_{3}}\sec\beta\label{eq:3h3},
\end{eqnarray}
where the terms $\chi_{3}$ is
\begin{eqnarray}
\chi_{3}=\sqrt{-2\frac{n_{2}}{n_{3}}(n_{2}+n_{3})\Omega-n_{1}\frac{n_{2}}{n_{3}}(n-1)c^{2}}.\label{eq:p2}
\end{eqnarray}
Note here that we have chosen $H_{2}$ is always greater than $H_{3}$.
Finally, using Eqs.~(\ref{eq:sumH}) and (\ref{eq:3h3}), one can also obtain the Hubble parameter $H_{3}$
\begin{eqnarray}
H_{2}=-\frac{\chi_{1}}{n-1}\tan\beta+\frac{\chi_{2}-n_{1}c}{n_{2}+n_{3}}\sec\beta\label{eq:3h2},
\end{eqnarray}
where the term $\chi_{2}$ is
\begin{eqnarray}
\chi_{2}=\sqrt{-2\frac{n_{3}}{n_{2}}(n_{2}+n_{3})\Omega-n_{1}\frac{n_{3}}{n_{2}}(n-1)c^{2}}\label{eq:p1}.
\end{eqnarray}
It is easy to see that the integration constant $c$ is constrained by Eqs.~(\ref{eq:p2}) and ~(\ref{eq:p1}).
The constant $c$ must be set to a value less than $c_{max}$ to guarantee that the value in the root is greater than zero, which yields
\begin{eqnarray}
c\leq\sqrt{\frac{-2\Omega(n_{2}+n_{3})}{n_{1}(n-1)}}\equiv c_{max},
\end{eqnarray}
where $c_{max}$ is the maximum value of $c$.
It is evident from Eqs.~(\ref{eq:3h3}), (\ref{eq:p2}), (\ref{eq:3h2}) and (\ref{eq:p1}) that, if $c=c_{max}$, $H_{2}$ is equal to $H_{3}$.
For convenience, we define a parameter $d$ satisfying
\begin{eqnarray}
c=dc_{max},
\end{eqnarray}
where the value of $d$ is between $0$ and $1$.
First, we investigate the effect of parameter $d$ on the Hubble parameters $H$.
We choose $n_{1}=3$ which is most in line with the presently observed three dimensional (3D) space.
Further setting $n_{2}=1$ and $n_{3}=1$,
we plot the Hubble parameters $H_{1}$, $H_{2}$ and $H_{3}$ as a function of $\beta$ in Fig.~\ref{fig:1} with $d=0$, $d=0.5$ and $d=1$ respectively.
In Fig.~\ref{fig:1}(a)-(c),
the three curves having same type (color) corresponds to three different value of $d$ respectively.
In Fig.~\ref{fig:1}(a), we plot the Hubble parameter $H_{1}$ as a function of $\beta$.
When the parameter $d=0$ and $d=0.5$, the Hubble parameters $H_{1}$ monotonically decrease as $\beta$.
And when $d>\sqrt{n_{1}/[(n_{2}+n_{3})(n-2)]}\equiv d_{min}=\sqrt{6}/4\approx0.612$ for $n_{1}=3$, $n_{2}=1$ and $n_{3}=1$, the Hubble parameter $H_{1}$ has a minimum
\begin{eqnarray}
H_{1min}=\frac{\sqrt{c^2(n-1)^2-\chi_{1}^{2}}}{n-1},
\end{eqnarray}
when the term $\beta$ in Eq.~(\ref{eq:3h1}) takes the form
\begin{eqnarray}
\beta_{min}=\arcsin[\frac{\chi_{1}}{c(n-1)}].
\end{eqnarray}
If $\beta\leq\beta_{min}$, the Hubble parameters $H_{1}$ is a monotonically decreasing function with time
which is not in accordance with the cosmological observations and experiments.
As can be seen easily from Fig.~\ref{fig:1}(a), the the Hubble parameters $H_{1}$ has a minimum when $d=d_{min}\approx0.612$.
This case may be consistent with the present observations because $H_{1}$ tends to a constant near the minimum,
which can lead to accelerated expansion without the contribution of dark energy (or an inflaton field).

\begin{figure}
\includegraphics[scale=0.68]{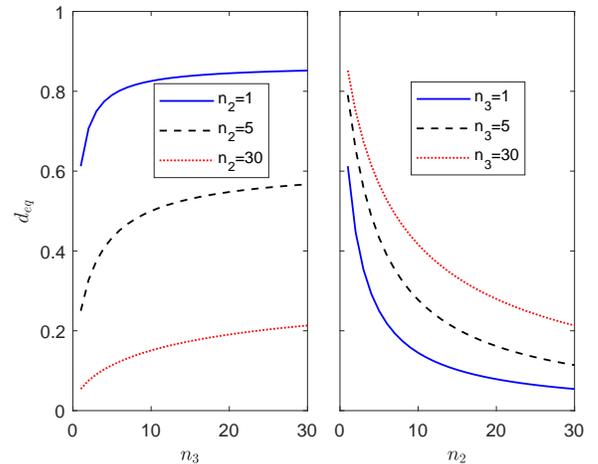}
\caption{\label{fig:2} The parameter $d_{eq}$ versus the number of extra dimensions $n_{2}$ and $n_{3}$ respectively.
The figure on the left is the curve of the parameter $d_{eq}$ with $n_{3}$ when $n_{2}=1, 5, 30$,
and the figure on the right depicts the evolution of $d_{eq}$ with $n_{2}$ when $n_{3}=1, 5, 30$.}
\end{figure}

\begin{figure*}
\includegraphics[scale=0.76]{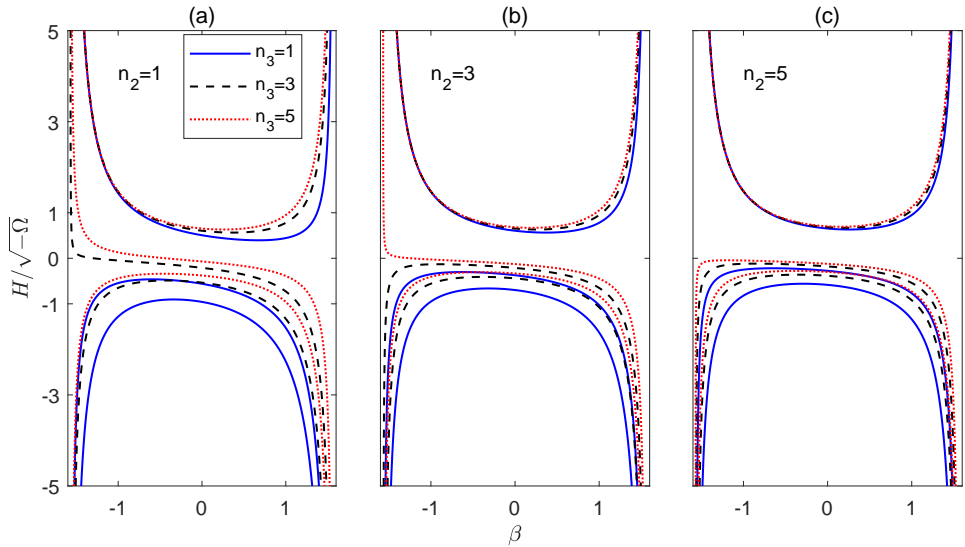}
\caption{\label{fig:3} The Hubble parameters $H_{1}$ ($n_{1}=3$), $H_{2}$ and $H_{3}$ when $d=0.98$.
(a) The Hubble parameters with $n_{3}=1, 3, 5$ correspond to the solid (blue) curve, the dashed (black) curve and the dotted curve (red) respectively ($n_{2}=1$). (b)-(c) The case for $n_{2}=3$ and $n_{2}=5$ respectively.}
\end{figure*}

Combining Eqs.~(\ref{eq:chi1}), (\ref{eq:3h1}), (\ref{eq:3h2}) and (\ref{eq:p1}), we obtain $H_{1}=H_{2}$ if $c$ is satisfied by
\begin{eqnarray}
c=\sqrt{\frac{-2\Omega n_{3}}{(n-1)(n_{1}+n_{2})}}\equiv c_{eq}.
\end{eqnarray}
It is shown that $H_{2}$ tends to $H_{1}$ when $c\rightarrow c_{eq}$.
In the case $c=c_{eq}$, the parameter $d_{eq}$ is defined as
\begin{eqnarray}
d_{eq}=\frac{c_{eq}}{c_{max}}=\sqrt{\frac{n_{1}n_{3}}{(n_{1}+n_{2})(n_{2}+n_{3})}}.
\end{eqnarray}
For $n_{1}=3$, $n_{2}=1$ and $n_{3}=1$, $d_{eq}=\sqrt{6}/4\approx0.612$
Since the extra dimension in our universe is not observed presently,
the extra dimension should be very small,
which requires that the extra dimensions cannot be too large and the current scale of extra dimensions are still outside the observable range.
When $d$ is equal $d_{eq}$, the extra dimension Hubble parameter $H_{2}$ is converted to $H_{1}$.
This case is inconsistent with the presently observed 3D space.
The Hubble parameter $H_{2}$ of the extra dimensions is plotted in Fig.~\ref{fig:1}(b) with $d=0, 0.5, 1$.
It can be easily shown that the Hubble parameter $H_{2}$ has a minimum with $d=0$ and $0.5$,
which can lead to accelerated expansion of the extra dimensions.
We are not interested in this situation because that it is not in line with the observation.
However, the Hubble parameter $H_{2}$ is always negative with $d=1$,
which ensures that the extra dimensions always exceeds the observable range.
Note here the ratio of $d_{min}$ to $d_{eq}$ is
\begin{eqnarray}
\frac{d_{min}}{d_{eq}}=\sqrt{\frac{(n_{1}+n_{2})}{n_{3}(n_{1}+n_{2}+n_{3}-1)}}\leq 1,
\end{eqnarray}
where $d_{min}/d_{eq}=1$ if and only if $n_{3}=1$.
So we only consider the case $d\gg d_{eq}$ because that we obtain $d\geq d_{min}$ when $d\geq d_{eq}$.

In Fig.~\ref{fig:2}, we have plotted the parameter $d_{eq}$ versus the number of extra dimensions $n_{2}$ and $n_{3}$ respectively.
The figure on the left is the curve of the parameter $d_{eq}$ with $n_{3}$ when $n_{2}=1$, $5$, and $30$.
It is shown that $d_{eq}$ monotonically increases with $n_{3}$.
The parameter $d_{eq}$ tend to $\sqrt{n_{1}/(n_{1}+n_{2})}$ in the $n_{3}\rightarrow\infty$ limit.
In particular, $d_{eq}\rightarrow\sqrt{3}/2\approx0.866$ for $n_{2}=1$.
The figure on the right depicts the evolution of $d_{eq}$ with $n_{2}$ when $n_{3}=1$, $5$, and $30$.
In this case, $d_{eq}$ monotonically decrease with $n_{2}$.
The parameter $d_{eq}\rightarrow0$ in the $n_{2}\rightarrow\infty$ limit.
To be consistent with observation,
we should set $d$ to be greater than $0.866$ and closer to $1$.
Further we set the constant $d=0.98$ in the following.

In Fig.~\ref{fig:3}(a)-(c), we plot the Hubble parameters $H_{1}$, $H_{2}$ and $H_{3}$ as a function of $\beta$ with $n_{1}=3$ when $d=0.98$.
From top to bottom,
the three curves having same type (color) corresponds to $H_{1}$, $H_{2}$ and $H_{3}$ respectively for fixed value of $n_{2}$ and $n_{3}$.
In Fig.~\ref{fig:3}(a), we plot the Hubble parameter as a function of $\beta$ at $n_{2}=1$.
From top to bottom, the three solid curves are correspond to $H_{1}$, $H_{2}$ and $H_{3}$ at $n_{3}=1$.
The three dashed curves ($n_{3}=3$) and the three dotted curves ($n_{3}=5$) are similar to the above case.
The Hubble parameter of the extra dimensions are always negative at $n_{3}=1$.
With the increase of $n_{3}$, the coordinate of the minimum value of $H_{1}$ tends to $\beta=0$.
Meanwhile, $H_{2}$ changes from positive to negative in the region near $\beta\sim-1.5$
and $H_{3}$ is closer to zero.
In Fig.~\ref{fig:3}(b) and (c), the Hubble parameters as a function of $\beta$ are shown with $n_{2}=3$ and $n_{2}=5$.
$H_{2}$ is always negative with $n_{2}=5$ which is similar to the situation in Fig.~(1) of Ref.~\cite{Kang2}.
$H_{1}$ has a lower bound $H_{1min}=\sqrt{0.98^{2}c_{max}^{2}-\chi_{1}^{2}/(n-1)^{2}}$ when $\beta_{min}=\arcsin\{\eta_{1}/[0.98(n-1)c_{max}]\}$.
In the region near $\beta_{min}$, we get $\Omega_{eff}>0$ is of order $-\Omega$.
This situation is similar to the case with two different scale factors,
the negative induced cosmological constant $\Omega$ can be transformed into the positive effective cosmological constant $\Omega_{eff}$.
It tells us that the observed current cosmic acceleration is intrinsically an extra-dimensional phenomenon rather than dark energy.
The cosmological constant fine tuning problem can be solved by this extra-dimensional evolution.

From Eqs.~(\ref{eq:3h1}), (\ref{eq:3h3}) and (\ref{eq:3h2}) we can get the scale factors $a_{1}$, $a_{2}$ and $a_{3}$ of the form
\begin{eqnarray}
a_{1}=a_{10}|\cos\beta|^{\frac{1}{n-1}}|\sec\beta+\tan\beta|^{\frac{c}{\chi_{1}}},\\
a_{2}=a_{20}|\cos\beta|^{\frac{1}{n-1}}|\sec\beta+\tan\beta|^{\frac{\chi_{2}-n_{1}c}{\chi_{1}(n_{2}+n_{3})}},\\
a_{2}=a_{30}|\cos\beta|^{\frac{1}{n-1}}|\sec\beta+\tan\beta|^{-\frac{\chi_{3}+n_{1}c}{\chi_{1}(n_{2}+n_{3})}},
\end{eqnarray}
where $a_{10}$ , $a_{20}$ and $a_{30}$ are the scale factors when the brane forms.
Further, the volume of the visible brane is obtain by
\begin{eqnarray}
V_{b}=a_{1}^{n_{1}}a_{2}^{n_{2}}a_{3}^{n_{3}}=a_{10}^{n_{1}}a_{20}^{n_{2}}a_{30}^{n_{3}}\cos\beta,
\end{eqnarray}
When the brane is just forming, there is no particular reason to make the scale factor different,
so we choose $a_{10}=a_{20}=a_{30}$.
If the initial scale of the brane is of order $10^{35}$ in planck unit,
and consider the presently observed scale of our universe (about of order $10^{61}$),
we obtain the scale of extra dimension is at least of order $10^{22}$ with $n_{2}=n_{3}=3$.
It is shown that the scale of extra dimension should be much larger than Planck length.
This ensures that physics are still valid in the evolution of the visible brane.
We obtain $\theta_{0}$ is close to $-\pi/2$ if one want obtain a sufficiently small initial scale.
So in the region of $\theta_{0}+\pi/2\ll\eta_{1}t\ll\pi/2$, the Hubble parameters $H_{1}$ is of the form:
\begin{eqnarray}
H_{1}&\simeq&[\frac{c}{\chi_{1}}+\frac{1}{n-1}]\frac{1}{t} \nonumber\\
&=&\frac{1}{n-1}[1+d\sqrt{\frac{(n_{2}+n_{3})(n-2)}{n_{1}}}]\frac{1}{t}.
\end{eqnarray}
When $n_{2}=n_{3}=1$ and $d=0.98$, $H_{1}$ is about $0.52t$ which is as similar as the radiation dominating eras.
In the limit $n_{2}\rightarrow\infty$ (or $n_{3}\rightarrow\infty$), $H_{1}\simeq\sqrt{3}d/3t$.

\section{Summary and Conclusion}
In conclusion, we investigate a $(n+1)$-d generalized Randall-Sundrum model with an anisotropic metric which has three different scale factors.
In this model, one obtain the positive effective cosmological constant $\Omega_{eff}\sim10^{-124}$ (in Planck unit) which only need a solution $kr\simeq50-80$ without fine tuning.
This is consistent with the case with two different scale factors.

In this model, the Hubble parameters $H_{2}$ tends to $H_{1}$ when the integration constant $d$ tends to $d_{eq}$.
It is indicate that the Hubble parameter of observable dimensions is related to the value of the integral constant $d$.
For convenience, here we have selected the Hubble parameter $H_{1}$ to show the observable dimensions.
To be consistent with observation,
we should set $d$ to be greater than $0.866$ and closer to $1$.
Further setting the constant $d=0.98$,
we obtain that $H_{1}$ has a lower bound $H_{1min}=\sqrt{0.98^{2}c_{max}^{2}-\chi_{1}^{2}/(n-1)^{2}}$ when $\beta_{min}=\arcsin\{\eta_{1}/[0.98(n-1)c_{max}]\}$.
Meanwhile, the scale of extra dimension is smaller than the observed scale but greater than the Planck length.
This demonstrates that the observed current cosmic acceleration is caused by the extra-dimensional evolution rather than dark energy (or an inflaton field).

\begin{acknowledgments}
We wish to acknowledge the supported by the Key Program of National Natural Science Foundation of China (under Grant No. 11535005),
the National Natural Science Foundation of China (under Grant No. 11647087),
Foundation for Young and Yiddle-Aged Teachers Basic Ability Improvement in Guangxi Universities (under Grant No. 2018KY0326)
Special Foundation for Science and Technology Base and Talents in Guangxi (under Grant No. 2018AD19310)
China Postdoctoral Science Foundation funded project (under Grant No. 2019M651750),
Open project of state key laboratory of solid state microstructure physics (under Grant No. M31037),
the Natural Science Foundation of Yangzhou Polytechnic Institute (under Grant No. 201917),
and the Natural Science Foundation of Changzhou Institute of Technology (Grant No. YN1509).
\end{acknowledgments}

\end{document}